\documentclass[12pt]{article}

\usepackage{amssymb}
\usepackage{amsmath}
\usepackage{amsfonts}
\usepackage{graphicx}
\usepackage{subfigure}
\usepackage{color}
\usepackage{dcolumn}
\usepackage{hyperref}
\numberwithin{equation}{section}
\newcommand{\be}{\begin{equation}}
\newcommand{\ee}{\end{equation}}
\newcommand{\bea}{\begin{eqnarray}}
\newcommand{\eea}{\end{eqnarray}}

\def\a{\alpha}      
       
  \def\G{\Gamma}

\def\l{\lambda}

\def\s{\sigma}

\newcommand{\eq}[1]{(\ref{#1})}

\begin{document}

\begin{titlepage}
\thispagestyle{empty}


\vspace{2cm}

\begin{center}
\font\titlerm=cmr10 scaled\magstep4 \font\titlei=cmmi10
scaled\magstep4 \font\titleis=cmmi7 scaled\magstep4 {
\Large{\textbf{The Imaginary Potential and Thermal Width of Moving
Quarkonium from Holography}
\\}}
\vspace{1.5cm} \noindent{{
Kazem Bitaghsir
Fadafan$^{}$\footnote{e-mail:bitaghsir@shahroodut.ac.ir }, Seyed
Kamal Tabatabaei$^{}$\footnote{e-mail:k.tabatabaei67@yahoo.com }
}}\\
\vspace{0.8cm}

{\it ${}$Physics Department, Shahrood University, Shahrood, Iran\\}

\vspace*{.4cm}

\end{center}

\vskip 2em

\begin{abstract}
We study the effect of finite 't Hooft coupling corrections on the
imaginary potential and the thermal width of a moving heavy
quarkonium from the AdS/CFT correspondence. To study these
corrections, we consider
 $\mathcal{R}^4$ terms and Gauss$-$Bonnet gravity. We conclude that
the imaginary potential of a moving or static heavy quarkonium
starts to be generated for smaller distances of quark and antiquark.
Similar to the case of static heavy quarkonium, it is shown that by
considering the corrections the thermal width becomes effectively
smaller. The results are compared with analogous calculations in a
weakly coupled plasma.

\end{abstract}

\end{titlepage}

\tableofcontents
\section{Introduction}
In heavy ion collisions at the LHC and RHIC, heavy quark related
observables are becoming increasingly important \cite{Laine:2011xr}.
In these collisions, one of the main experimental signatures of the
formation of a strongly coupled quark$-$gluon plasma (QGP) is
melting of quarkonium systems, like $J/\psi$ and excited states, in
the medium. In this case, the main mechanism responsible for this
suppression is color screening \cite{Matsui:1986dk}. Recent studies
suggest a more important reason than screening, which is the
existence of an imaginary part of the potential
\cite{Lain1,Escobedo:2014gwa}. Then the thermal width of these
systems is an important subject in QGP. In the effective field
theory framework, thermal decay widths have been studied in
\cite{Brambilla:2008cx,Brambilla:2010vq}. It was shown that at
leading order, two different mechanisms contribute to the decay
width, namely Landau damping and singlet-to-octet thermal breakup
\cite{Brambilla:2011sg}. The analytic estimate of the imaginary part
of the binding energy and the resultant decay width were studied in
\cite{Dumitru:2010id}. The peak position and its width in the
spectral function of heavy quarkonium can be translated into the
real and imaginary part of the potential \cite{Rothkopf:2011db}. The
effect of the imaginary part of the potential on the thermal widths
of the states in both isotropic and anisotropic plasmas has been
studied in \cite{Margotta:2011ta}.

In heavy ion collisions, most of quarkonium observed experimentally
are moving through the medium with relativistic velocities. They
have finite probability to survive, even at infinitely high
temperature. Two mechanisms of charmonium attenuation by considering
final state interactions have been studied in
\cite{Kopeliovich:2014una}. The bottomium suppression in PbPb
collisions at LHC energies is studied in \cite{Nendzig:2014qka}.
Study of heavy quarkonium moving in a quark-gluon plasma from
effective field theory techniques has been done in
\cite{Escobedo:2013tca, Song:2007gm}. It is shown that in the regime
relevant for dissociation and for very large velocities the thermal
width decreases with increasing velocity of quarkonium
\cite{Escobedo:2013tca}.

A new method for studying different aspects of QGP is the AdS/CFT
correspondence \cite{CasalderreySolana:2011us}. This method has
yielded many important insights into the dynamics of strongly
coupled gauge theories. It has been used to study hydrodynamical
transport quantities at equilibrium and real$-$time processes in
non-equilibrium \cite{DeWolfe:2013cua}. Methods based on AdS/CFT
relate gravity in $AdS_5$ space to the conformal field theory on the
four-dimensional boundary \cite{Witten:1998qj}. It was shown that an
$AdS$ space with a black brane is dual to a conformal field theory
at finite temperature \cite{Witten:1998zw}.

We have studied the imaginary potential and the thermal width of
static quarkonium from the AdS/CFT correspondence in
\cite{Fadafan:2013coa}. These quantities initially studied in
\cite{Noronha:2009da}. In this approach, the thermal width of heavy
quarkonium states originates from the effect of thermal fluctuations
due to the interactions between the heavy quarks and the strongly
coupled medium. This method was revisited in \cite{Finazzo:2013rqy}
and general conditions for the existence of an imaginary part for
the heavy quark potential were obtained. In the context of AdS/CFT,
there are other approaches which can lead to a complex static
potential \cite{Albacete:2008dz,Hayata:2012rw}. The case of
anisotropic plasma has been studied in
\cite{Dimitris1,Fadafan:2013bva} and imaginary potential formula in
a general curved background was obtained. It was also shown that the
thermal width is decreased in presence of anisotropy and bigger
decrease happens along the transverse plane. With using a probe
D7-brane in the dual gravity theory, thermal width is investigated
in \cite{Ali-Akbari:2014gia}. The effects of deformation parameter
on thermal width of moving quarkonium has been studied in
\cite{Sadeghi:2014zya}.

The imaginary potential and the thermal width of a moving heavy
quark$-$antiquark pair in an isotropic plasma, depend on the angle
$\theta$ between the direction of the pair and the velocity of the
wind $v$ . Then it would be possible to consider different
alignments for quark$-$antiquark pair with respect to the plasma
wind. They are parallel ($\theta=0$), transverse ($\theta=\pi/2$) or
arbitrary direction to the wind ($\theta$). By increasing the angle
from zero to $\pi/2$ the imaginary potential and the thermal width
for a fixed velocity is changing. We will examine this idea in the
presence of higher derivative corrections. Including the higher
derivative corrections would be important from a phenomenology point
of view. Two extreme cases, transverse and parallel to the wind have
been studied in \cite{Ali-Akbari:2014vpa}. The case where the axis
of the moving pair has an arbitrary orientation with respect to the
wind has been done in \cite{Finazzo:2014rca}. It should be noticed
that the extreme cases are much simpler than arbitrary case to
study. In this article, we have studied all cases.

Using the AdS/CFT correspondence, we have studied melting of a heavy
meson like $J/ \psi$ and excited states like $\chi_c$ and $\psi'$ in
the quark medium in \cite{Fadafan:2012qy}. It was shown that the
excited states melt at higher temperatures. Also heavy quarks in the
presence of higher derivative corrections have been studied in
\cite{Noronha:2009ia,Fadafan:2011gm}. In this paper, we continue our
study in \cite{Fadafan:2013coa} to a moving quarkonium. Finite 't
Hooft coupling corrections have been considered. These corrections
correspond to $\mathcal{R}^4$ corrections and Gauss$-$Bonnet terms,
respectively. An understanding of how the imaginary part of the
potential and the thermal width of heavy quarkonium are affected by
these corrections may be essential for theoretical predictions.
Actually, to study more realistic models we added higher derivative
Gauss$-$Bonnet corrections. They are the leading $1/N_c$ corrections
in the presence of a D7-brane. It has been shown that these type of
corrections can increase the nuclear modification factor $R_{AA}$
\cite{Ficnar:2013qxa,Ficnar:2012yu}.

We compare our results with static case and conclude that the
imaginary potential of a moving or static heavy quarkonium starts to
be generated for smaller distances of quark and antiquark. Also
similar to the case of static heavy quarkonium, it is shown that by
considering these corrections the thermal width becomes effectively
smaller. The same results also can be found in calculations in a
 weakly coupled plasma \cite{Escobedo:2013tca}.

This paper is organized as follows. In the next section, we will
present the finite coupling corrections. We give the general
expressions to study the imaginary potential and the thermal width
for ($\theta=\pi/2$) in Sect. 3. For ($\theta=0$) and arbitrary
angle ($\theta$), we address the relevant papers and concern mainly
on the results. Also in this section, we show the effect of the
corrections in the different plots. The thermal width behavior has
been explored in Sect. 4. We discuss the limitations of the method,
too. In the last section we summarize our results. %

\section{On finite coupling corrections}
One may consider the general gravity as follows:%
\be
 ds^2=G_{tt}dt^2+G_{xx}dx_i^2+G_{uu}du^2,\label{general-background}
\ee%
here the metric elements are functions of the radial distance $u$
and $x_i=x,y,z$ are the boundary coordinates. In these coordinates,
the boundary is located at infinity.

From the AdS/CFT, the coupling which is denoted as 't Hooft coupling
$\lambda$ is related to the curvature radius of the $AdS_5$ and
$S_5$ $(R)$, and the string tension $(\frac{1}{2\pi \alpha'})$ by
$\sqrt{\lambda}=\frac{R^2}{\alpha'}$. A general result of the
AdS/CFT correspondence states that the effects of finite but large
$\lambda$ coupling in the boundary field theory are captured by
adding higher derivative interactions in the corresponding
gravitational action. In our study, the corrections to the
AdS-Schwarzschild black brane are $\mathcal{R}^4$ and
$\mathcal{R}^2$ corrections.

\begin{itemize}%
\item{$\mathcal{R}^4$ corrections:}%
\end{itemize}%
Since the $AdS/CFT$ correspondence refers to complete string theory,
one should consider the string corrections to the 10D supergravity
action. The first correction occurs at order $(\alpha')^3$
\cite{alpha2}. In the extremal $AdS_5\times S^5$ it is clear that
the metric does not change \cite{Banks}; conversely this is no
longer true in the non-extremal case. Corrections in inverse 't
Hooft coupling $1/\lambda$, which correspond to $\alpha^{\prime}$
corrections on the string theory side, were found in \cite{alpha2}.
The functions of the $\alpha^{\prime}$-corrected metric are given by
\cite{alpha1}
\begin{eqnarray}\label{correctedmetric}
G_{tt}&=&-u^2(1-w^{-4})T(w),\nonumber\\
G_{xx}&=&u^2 X(w),\nonumber\\
G_{uu}&=&u^{-2}(1-w^{-4})^{-1} U(w),
\end{eqnarray}
where%
\begin{eqnarray}
T(w)&=&1-k\bigg(75w^{-4}+\frac{1225}{16}w^{-8}-\frac{695}{16}w^{-12}\bigg)+\dots ,\nonumber\\
X(w)&=&1-\frac{25k}{16}w^{-8}(1+w^{-4})+\dots,\nonumber\\
U(w)&=&1+k\bigg(75w^{-4}+\frac{1175}{16}w^{-8}-\frac{4585}{16}w^{-12}\bigg)+\dots,\
\end{eqnarray}
and $w=\frac{u}{u_h}$. There is an event horizon at $u=u_h$ and the
geometry is asymptotically $AdS$ at large $u$ with a radius of
curvature $R=1$. The expansion parameter $k$ can be expressed in
terms
of the inverse 't Hooft coupling as %
\be\label{k}%
 k=\frac{\zeta(3)}{8}\lambda^{-3/2}\sim 0.15\lambda^{-3/2}.
\ee %
The temperature is given by%
\be %
 T_{}=\frac{u_h}{\pi R^2 (1-k)}.
\ee %

\begin{itemize}%
\item{$\mathcal{R}^2$ corrections:}%
\end{itemize}%

In five dimensions, we consider the theory of gravity with quadratic
powers of curvature as Gauss$-$Bonnet (GB) theory. The exact
solutions and thermodynamic properties of the black brane in GB
gravity are
discussed in \cite{Cai:2001dz,Nojiri:2001aj,Nojiri:2002qn}. The metric functions are given by%
\begin{equation}
G_{tt}=-N \,u^2\, h(u),\,\,\,\,\,\, G_{uu}=\frac{1}{u^2
h(u)},\,\,\,\,\, G_{xx}=G_{yy}=G_{zz}=u^2\label{GBmetric},
\end{equation}
where
\begin{equation}
h(u)= \frac{1}{2\lambda_{GB}}\left[ 1-\sqrt{1-4 \lambda_{GB}\left(
1-\frac{u_h^4}{u^4} \right)}\right].
\end{equation}
In (\ref{GBmetric}), $N= \frac{1}{2}\left(
 1+\sqrt{1-4 \lambda_{GB}} \right)$ which is an arbitrary constant that specifies
the speed of light of the boundary gauge theory and we choose it to
be unity. Beyond $\l_{GB}<1/4$ there is no vacuum AdS solution and
one cannot have a conformal field theory at the boundary. Causality
leads to new bounds for the value of the Gauss$-$Bonnet coupling
constant as $-7/36<\l_{GB}<9/100$ \cite{Brigante008gz}. In our
method for computing the thermal width, we find that only positive
values of $\l_{GB}$ is valid. The temperature also is given by
\begin{equation}
 T_{}=\sqrt{N}\,\frac{u_h}{\pi R^2}.
\end{equation}
Also the 't Hooft coupling of the dual strongly coupled CFT is
$\l=\frac{N^2 R^4}{\a'^2}$ .

\section{Finding the imaginary potential }
In our study, we assume that the quark$-$antiquark pair is moving
with rapidity $\eta$. One should notice that in our reference frame
the plasma is at rest. Then we boost the frame in the $x_3$
direction with rapidity $\eta$ so that $dt' = dt' \cosh \eta - dx'_3
\sinh \eta$ and $dx_3 = -dt' \sinh \eta + dx'_3 \cosh \eta.$ After
dropping the primes, the metric
becomes%
\begin{align}
\label{metricboost} ds^2 = & -\left(|G_{tt}| \cosh^2 \eta - G_{xx}
\sinh^2 \eta\right) dt^2 + \left(G_{xx} \cosh^2 \eta - |G_{tt}|
\sinh^2 \eta\right) dx_3^2  \nonumber \\ & -2\,\sinh \eta \, \cosh
\eta\left(G_{xx}-|G_{tt}|\right)  \, dt \, dx_3 + G_{xx}
(dx_1^2+dx_2^2) + G_{uu} du^2,
\end{align}%

We define new metric functions as follows:%
\begin{subequations}
\label{WVtilde}
\begin{align}
\tilde{W} (u) \equiv W(u) \cosh^2 \eta - N(u) \sinh^2 \eta \\
\tilde{V} (u) \equiv V(u) \cosh^2 \eta - P(u) \sinh^2 \eta
\end{align}
\end{subequations}
where $V(u) \equiv -|G_{tt}|G_{xx},\,\,\,W(u)\equiv
-|G_{tt}|G_{uu},\,\,P(u)=G_{xx}^2$ and $N(u)=G_{xx}G_{uu}.$

Now one should use the usual orthogonal Wilson loop, which
corresponds to the heavy Q\={Q} pair.%
\subsection{Transverse to the wind ($\theta=\pi/2$)}
Assuming the system to be aligned perpendicularly to the wind, in
the $x_1$
direction %
\be t=\tau,\quad x_1=\sigma,\quad
u=u(\sigma)~,\label{static-gauge} \ee%
The distance between $Q\bar{Q}$ pair depends on the velocity and
angle, $L(\pi/2,\eta)$, we call it $L$.  The quarks are located at
$x_1=-\frac{L}{2}$ and $x_1=+\frac{L}{2}$. One finds the following
generic formulas for the heavy meson. In these formulas $u_*\equiv
u(x=0)$ is the deepest point of the U-shaped string.

\begin{itemize}
\item{The distance between quark and antiquark, $L$, is given by \be\label{staticL1}
L(\pi/2,\eta)=2\,\int_{u_*}^{\infty}du~\left[
\frac{\tilde{V}(u)}{\tilde{W}(u)}\left(\frac{\tilde{V}(u)}{\tilde{V}(u_*)}-1\right)\right]
^{-\frac{1}{2}}.  \ee }

\item{The real part of heavy quark potential,
$Re V_{Q\bar{Q}}(\pi/2,\eta)$ is as follows: \bea Re V_{Q\bar{Q}}=
\frac{1}{\pi \a'}\left[\int_{u_*}^{\infty}du~\left(
\left(\frac{1}{\tilde{W}(u)}-\frac{\tilde{V}(u_*)}{\tilde{V}(u)\tilde{W}(u)}\right)^{-1/2}-
\sqrt{\tilde{W}_0(u)}\right)-\int_{u_h}^{u_*}du~\sqrt{\tilde{W}_0(u)}
\right],\nonumber\\\label{staticV} \eea   here
$\tilde{W}_0(u)=\tilde{W}(u\rightarrow \infty)$. }

\item{The imaginary part of the potential is negative and it is given by
\be \text{Im} V_{Q\bar{Q}}(\pi/2,\eta)=
-\frac{1}{2\sqrt{2}\alpha'}\left[\frac{\tilde{V}'(u_*)}{2\tilde{V}''(u_*)}-\frac{\tilde{V}(u_*)}{\tilde{V}'(u_*)}
\right]~\sqrt{\tilde{W}(u_*)}~.  \label{ImV} \ee%
The derivatives are with respect to $u$. For the static quarkonium
and special case of $ W(u) = 1$, this formula reduces to the case of
an isotropic plasma \cite{Noronha:2009da}, while $ W(u)\neq 1$
corresponds to the anisotropic plasma
\cite{Dimitris1,Fadafan:2013bva} .}

\end{itemize}

These generic formulas give the related information of the moving
heavy quarkonium in terms of the metric elements of a background
\eq{general-background}.

To find $\text{Im}V$, one should express it in terms of the length
$L$ of the Wilson loop instead of $u_*$ using the equation
\eq{staticL1}. There is an important point for long U-shaped
strings, because it would be possible to add new configurations
\cite{Bak:2007fk}. Here we are interested mostly in distances $L
T<1$ and do not consider such configurations.

To begin with, we calculate the imaginary potential in the $N=4$
SYM. The metric
functions are %
\be
V(u)=\frac{u^4-u_h^4}{R^4},\,\,\,\,\,W(u)=1,\,\,\,P(u)=\frac{u^4}{R^4}
,\,\,\,N(u)=\left(1-\frac{u_h^4}{u^4}\right)^{-1}.\ee%
where the horizon is located at $u_h$ and the temperature of the hot
plasma is given by $u_h=\pi R^2 T$. In this case we do not consider
any correction and find the imaginary
potential from \eq{ImV} to be %
\be \text{ImV}_{Q\bar{Q}}^{}(\pi/2,\eta)= -\frac{\pi
\sqrt{\lambda}\,T}{24 \sqrt{2} \xi}\,\sqrt{\frac{1-\xi^4
cosh^2\eta}{1-\xi^4}}\left(3\xi^4cosh^2\eta-1\right),\,\,\,\,\,\,\,
\xi=\frac{u_h}{u_*}.\label{ImV0}\ee%
The imaginary potential is negative; it implies that there is a
lower bound for the deepest point of the U-shaped string, i.e.
$\xi_{min}^2=\frac{1}{\sqrt{3} cosh^2\eta}$. This minimum value is
found by solving $\text{ImV}_{Q\bar{Q}}^{}=0$. The maximum value
$\xi_{max}$ occurs when the distance $L$ approaches the maximum
value.
\begin{figure}
\centerline{\includegraphics[width=3in]{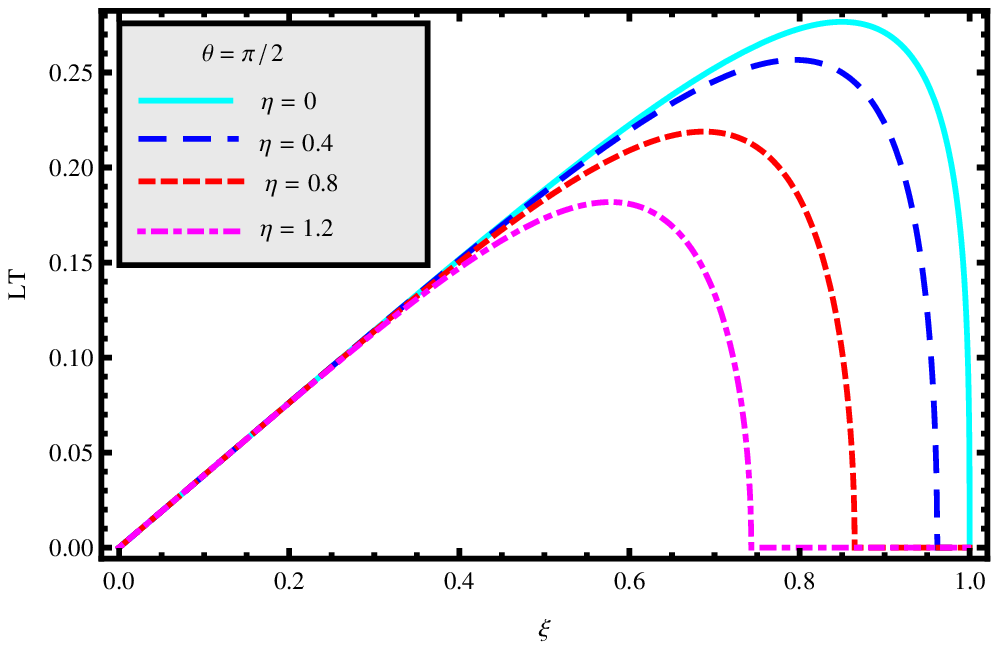}\,\,\includegraphics[width=3in]{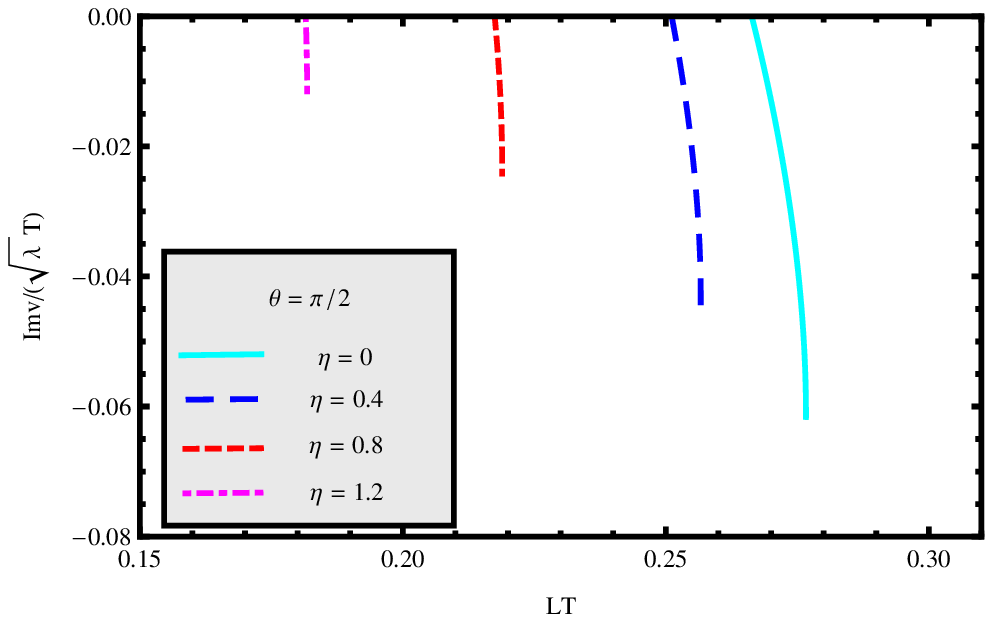}}%
\caption{ Left: The distance between quark and antiquark versus the
$\xi$ when the quarkonium is moving transverse to the wind. Right:
The imaginary potential for the valid values of $\xi$ or $L$.
}\label{N4Lxi90}
\end{figure}
The imaginary potential for accepted values of $\xi$ is shown in the
right plot of Fig. \ref{N4Lxi90}. It is clearly seen that the
imaginary potential starts to be generated at $L_{min}$ or
$\xi_{min}$ and increases in absolute value with $L$ until a value
$L_{max}$ or $\xi_{max}$. It should be noticed that for $\xi$ very
close to the horizon one should consider higher order corrections
\cite{Bak:2007fk}.%

\subsection{Parallel to the wind $(\theta=0)$}
Assuming the system to be aligned parallel to the wind, in the $x_3$
direction %
\be t=\tau,\quad x_3=\sigma,\quad
u=u(\sigma)~,\label{parallel static-gauge} \ee%
The distance between $Q\bar{Q}$ pair is called again $L$ and quarks
are located at $x_3=-\frac{L}{2}$ and $x_3=+\frac{L}{2}$. One finds
the generic formulas for finding $L$ and imaginary potential in
\cite{Ali-Akbari:2014vpa}. Here, we focus on the resultant plots.

\begin{figure}
\centerline{\includegraphics[width=3in]{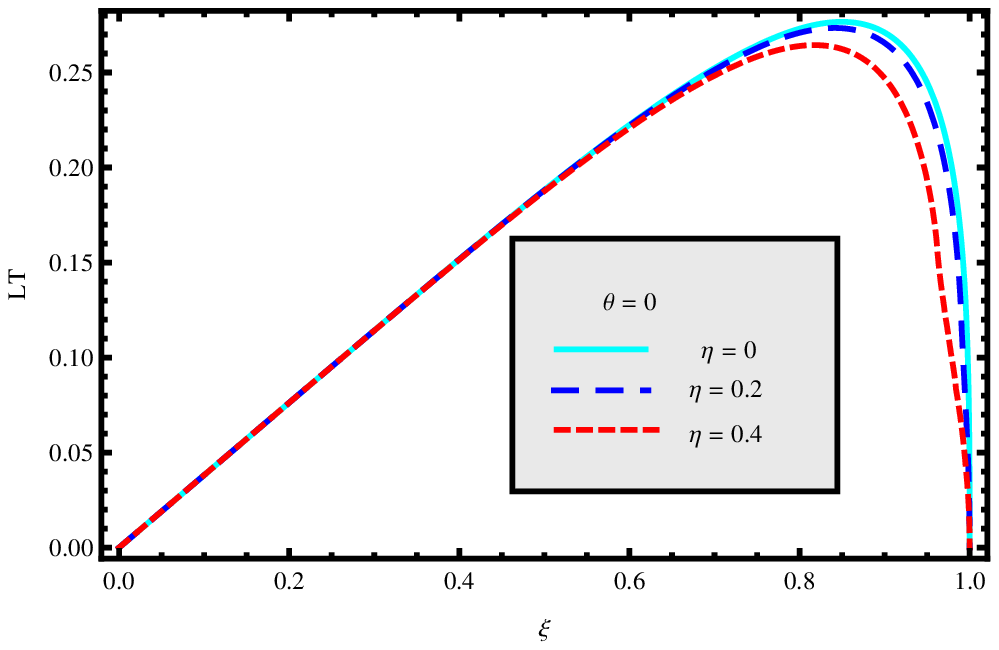}\,\,\includegraphics[width=3in]{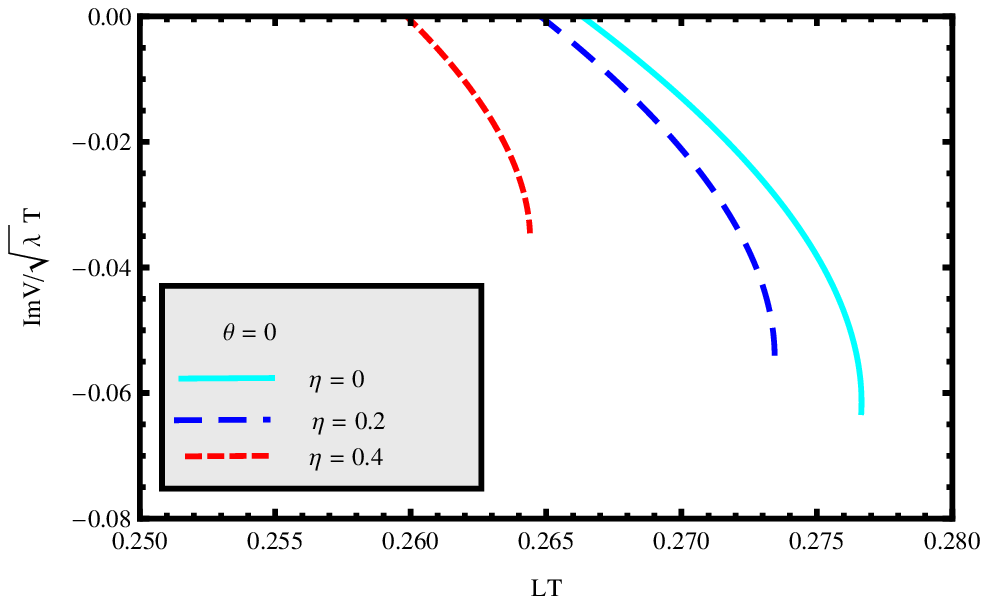}}%
\caption{ Left: The distance between quark and antiquark versus the
$\xi$ when the quarkonium is moving parallel to the wind. Right: The
imaginary potential for the accepted values of $\xi$ or $L$.
}\label{N4Lxi0}
\end{figure}

In the left plot of Fig. \ref{N4Lxi0}, the distance between quark
and antiquark versus the $\xi$ has been shown. The imaginary
potential also is given in the right plot of this figure. Comparing
the absolute value of imaginary potential in two extreme cases has
been done in \cite{Ali-Akbari:2014vpa,Finazzo:2014rca}. It is found
that stronger modification occurs when quarkonium is moving
orthogonal to the wind. This is similar to the screening length
where decreases as the quarkonium is moving transverse with respect
to the wind \cite{Liu:2006he}. One finds that by increasing rapidity
the imaginary potential happens for smaller LT. Thus, the magnitude
of the imaginary potential vanishes for ultra-relativistic moving
quarkonium. One important reason for this observation is coming from
the saddle point approximation. It was argued that this
approximation correspond to large velocity of moving quarkonium
\cite{Finazzo:2014rca}. Then one may discuss only the case of slowly
moving quarkonium.

\subsection{Arbitrary angle to the wind ($\theta$)}
Assuming the wind in the $x_3$ direction, we consider the quarkonium
to be aligned arbitrary to the wind as $\theta$ angle. Then the
parametrization
changes%
\be t=\tau,\quad x_1=\sigma, \quad x_3=x_3(\sigma),\quad
u=u(\sigma)~,\label{arbitrary} \ee%
The distance between $Q\bar{Q}$ pair is called again $L$ and the
projections of the system of quarks on the $x_1$ and $x_3$ direction
are of length $L\sin\theta$ and $L\cos\theta$, respectively. In this
case, the shape of the string is found by $x_3(\s)$ and $u(\s)$.
Especially, the $u(\s)$ is not a straight line. More details about
the shape of string and solutions are given in \cite{Liu:2006he}.
The velocity dependence of the potential also has been discussed.
One finds the generic formulas for finding the imaginary potential
in \cite{Finazzo:2014rca}. We have done the same calculations as
before. We found the distance between quark$-$antiquark ($Q\bar{Q}$)
versus the $\xi$ for different angles. Also the imaginary potential
for different angles has been studied. For example in the left plots
of Fig. \ref{ImVR4} and Fig. \ref{ImVGB}, the $\theta=\pi/3$ has
been considered.

One finally concludes the main results about the imaginary potential
as follows:
\begin{itemize}%
\item{The imaginary potential strongly depends on the velocity of the moving quarkonium and
the angle $\theta$.}%
\item{At fixed angle, by increasing the velocity the absolute value of the imaginary potential decreases. }
\item{At fixed velocity, by increasing the $\theta$ the absolute value of the imaginary potential increases. }
\item{The velocity of moving quarkonium has important effect when it
is moving transverse to the wind comparing with the parallel
motion.}
\end{itemize}%

\subsection{Finite coupling corrections}
\begin{itemize}%
\item{$\mathcal{R}^4$ corrections}
\end{itemize}%

\begin{figure}
\centerline{\includegraphics[width=3in]{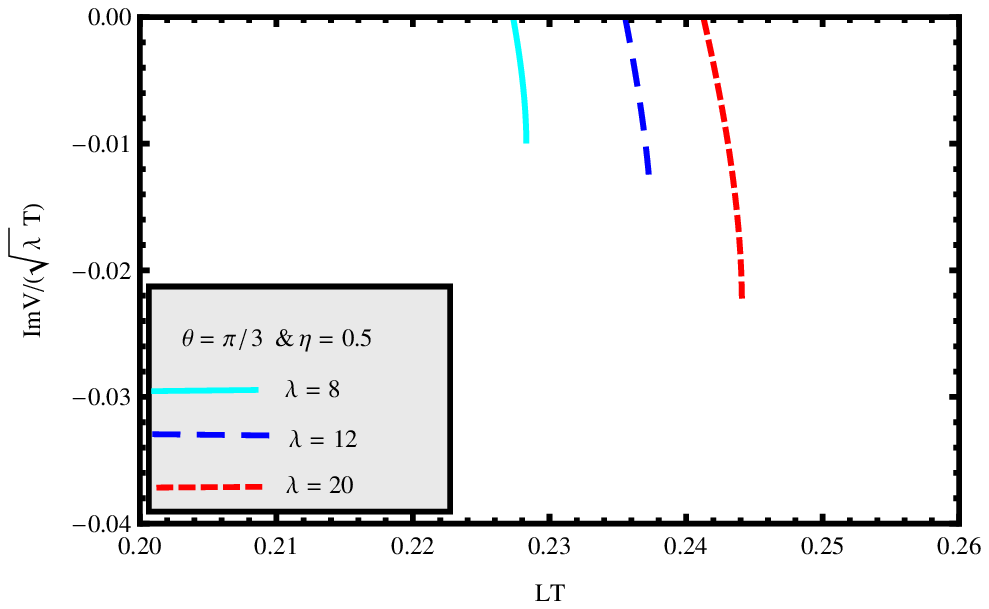}\,\includegraphics[width=3in]{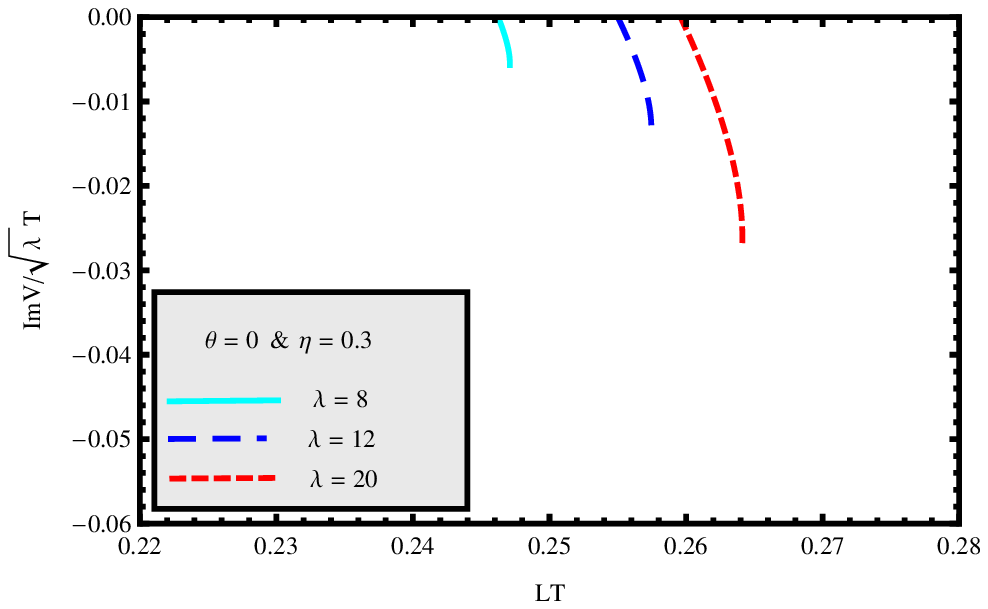}}%
\caption{ The imaginary potential versus $LT$ in the presence of
$\mathcal{R}^4$ corrections. Left: $(\theta=\pi/3, \eta=.5)$. Right:
$(\theta=0, \eta=.3)$. In each plot we assume $\lambda_{}$= 20, 12,
8. }\label{ImVR4}
\end{figure}

The distance between the quark and antiquark versus the $\xi$ for
different values of the coupling $\lambda_{}$ has been studied. We
find that decreasing $\lambda$ leads to decreasing of the maximum
value of the quark$-$antiquark distance. This behavior is the same
as the static case in Fig. 2 of \cite{Fadafan:2013coa}. The
imaginary potential at $(\theta=\pi/3, \eta=.5)$ and $(\theta=0,
\eta=.3)$ have been shown in the left and right plot of Fig.
\ref{ImVR4}, respectively. In each plot, we assume
$\lambda_{}=8,12,20$. As it was pointed out in the section 2, such
corrections decrease the 't Hooft coupling from infinity to finite
number. Thus, turning on $\mathcal{R}^4$ corrections leads to
generating the imaginary potential for smaller quark$-$antiquark
distances. This behavior is the same as the static quarkonium
\cite{Fadafan:2013coa}.

\begin{itemize}%
\item{Gauss$-$Bonnet corrections}
\end{itemize}

\begin{figure}
\centerline{\includegraphics[width=3in]{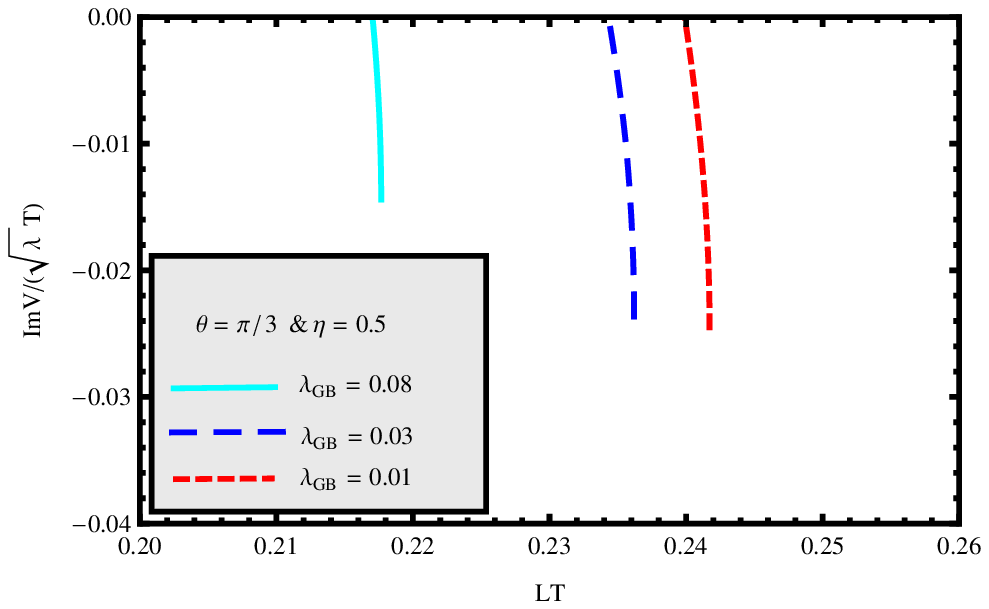}\,\,\includegraphics[width=3in]{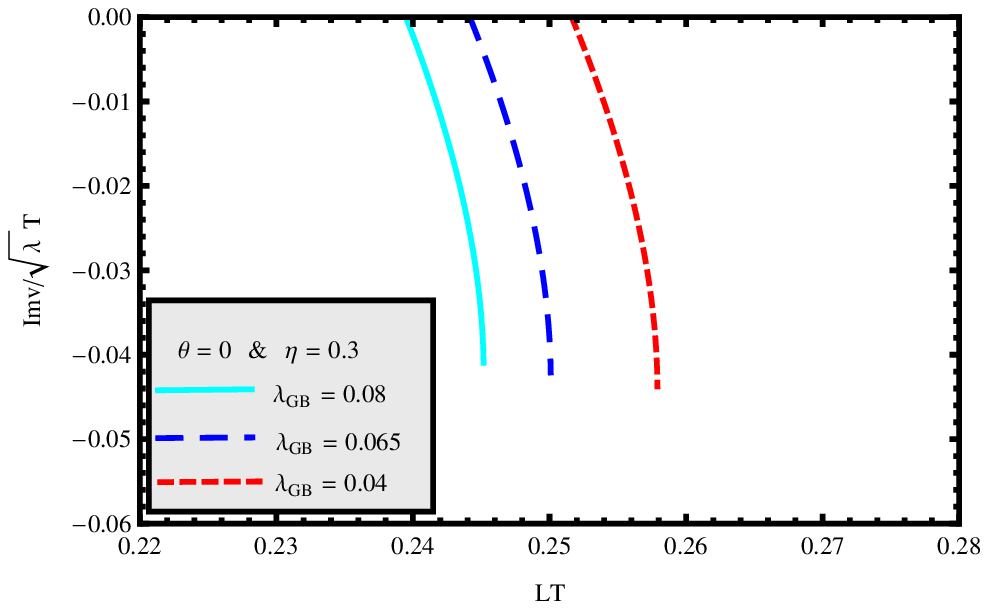}}%
\caption{The imaginary potential versus $LT$ in the presence of
Gauss$-$Bonnet corrections. Left: From left to right
$\lambda_{GB}$=0.08, 0.03, 0.01 and $(\theta=\pi/3, \eta=.5)$.
Right: From left to right $\lambda_{GB}$=0.08, 0.065, 0.04 and
$(\theta=0, \eta=.3)$. }\label{ImVGB}
\end{figure}

We find that only positive values of $\l_{GB}$ are valid in our
approach to produce the imaginary potential. The study of the
behavior of $L$ in terms of $\xi$ for different values of $\l_{GB}$
shows that by increasing the coupling constant the maximum value of
$L$ decreases. As in the static case, this behavior is not the same
as the $\mathcal{R}^4$ corrections \cite{Fadafan:2013coa}. The
imaginary potential at $(\theta=\pi/3, \eta=.5)$ and $(\theta=0,
\eta=.3)$ have been shown in the left and right plot of Fig.
\ref{ImVGB}, respectively. In each plot, we assume
$\lambda_{GB}$=0.04, 0.065, 0.08. The imaginary potential in the
Gauss$-$Bonnet gravity starts to be generated for smaller distances.
It means that by turning on the Gauss$-$Bonnet coupling, the
imaginary potential starts to be generated for smaller distances of
quark$-$antiquark. Also the absolute value of the imaginary
potential becomes smaller. The effect of higher derivative
corrections on the imaginary potential of a static quarkonium has
the same behavior \cite{Fadafan:2013coa}. This would be a general
feature of higher derivative corrections:
\begin{itemize}%
\item{The imaginary part of the potential of a moving or static heavy quarkonium
starts to be generated for smaller distances of quark$-$antiquark.}
\item{The absolute value of the imaginary part of the potential
of a moving or static heavy quarkonium becomes smaller.}
\end{itemize}%

We call the maximum distance of quark$-$antiquark where the
imaginary part of the potential starts as $LT_{max}$. To more
clarify the above general results, we plot $LT_{max}$ versus the
rapidity $\eta$ in Fig. \ref{L1Tma}. We see that $LT_{max}$
decreases monotonically with $\eta$. Such a behavior is the same as
the $LT_{min}$ for the onset of the imaginary part of the potential,
as shown in \cite{Finazzo:2014rca}. It is clearly seen that higher
derivative corrections decreases this quantity.

\begin{figure}
\centerline{\includegraphics[width=3in]{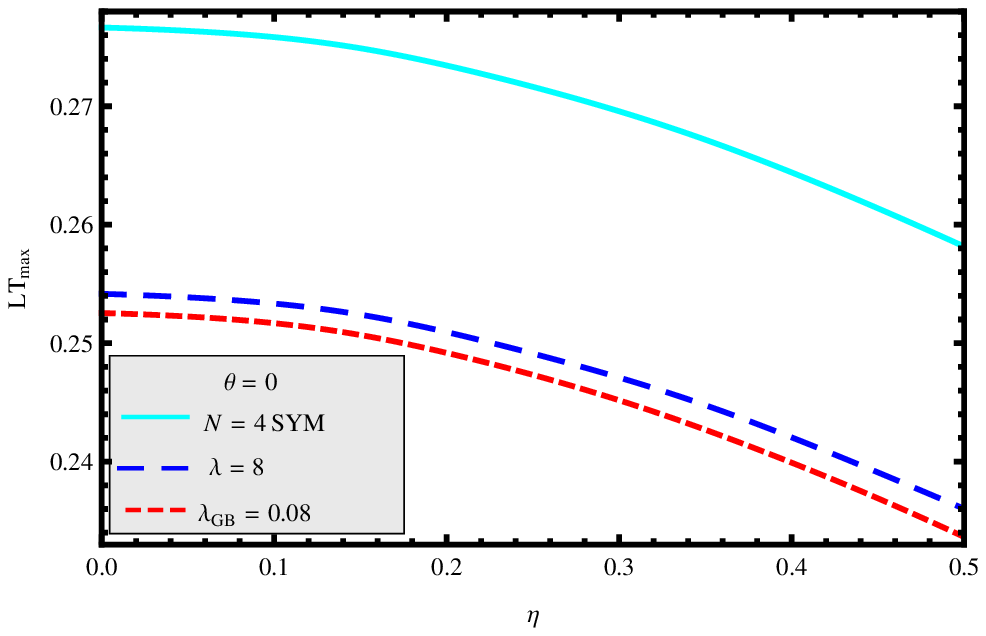}\,\,\includegraphics[width=3in]{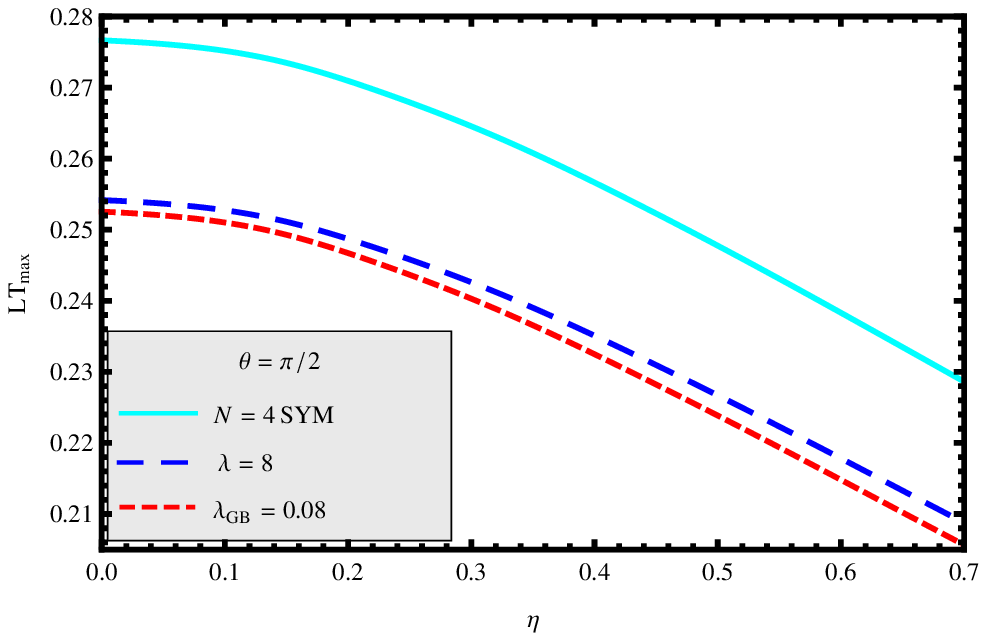}}%
\caption{$LT_{max}$ versus the rapidity for different corrections.
Left: Parallel to the wind $(\theta=0)$. Right: transverse to the
wind $(\theta=\pi/2)$. }
\end{figure}\label{L1Tma}

\section{Thermal width from holography}
In this section, we consider the moving quarkonium and compute the
corrections on the thermal width from holography. The static case
has been studied in \cite{Fadafan:2013bva}, it was also pointed out
that the imaginary potential for confining geometry is zero. To
calculate the thermal width of a heavy quarkonium like $\Upsilon$
meson, we use a first-order non-relativistic expansion,%
 \be
\Gamma=-<\psi|\text{ImV}_{Q\bar{Q}}|\psi>,\label{gamma} \ee %
also $|\psi>$ is the Coulombic wave function of the Coulomb
potential of the heavy quarkonium. Important points about finding
the Coulombic part in the presence of different corrections have
been discussed in \cite{Fadafan:2013bva}. By finding the imaginary
potential, the thermal width in the ground state of the energy
levels of the bound
state of heavy quarks with mass of $m_Q$ finds from \eq{gamma} as %
 \be \Gamma=-\frac{4}{a_0^3}\int_0^\infty L^2 d L e^{-2 L/ a_0}\text{ImV}_{Q\bar{Q}}^{}. \label{TWdefine}\ee
also the Bohr radius is defined as $a_0=2/m_Q A$. And one should
find $A$ from Columbic-like potential $-A/L$ \footnote{see
\cite{Fadafan:2013bva} for more details.}. Then for each velocity of
quarkonium the Bohr radius should also computed. We will use the
previous results to compute the thermal width.%

\subsection{Limitations of the method}
Now we explain the limitation to the calculation of the thermal
width from holography \cite{Finazzo:2013rqy,Fadafan:2013bva}. The
point is that the imaginary potential is defined in the region
$(L_{min}~,~L_{max})$ while we take the integral in \eq{TWdefine}
from zero to infinity. This valid region in the case of moving
quarkonium depends on the velocity. As it was shown, the absolute
value of $ImV$ vanishes by increasing the velocity. To compute the
width, one may consider the region $(L_{min}~,~\infty)$ by using a
reasonable extrapolation (straight line fitting) for imaginary
potential. This approach has been followed in \cite{Noronha:2009da}
for a static quarkonium and in \cite{Ali-Akbari:2014vpa} for a
moving quarkonium. When one considers the straight-line fitting for
$ImV$, we call it "approximate solution". More details on using this
method has been discussed also in \cite{Fadafan:2013bva}. One should
notice that the imaginary potential is not defined in the other
regions $(0~,~L_{min})$ and $(L_{max}~,~\infty)$. Then one may
consider the valid region $(L_{min}~,~L_{max})$ and compute the
width. In this case, we call the solution as "exact". As it was
explained in \cite{Finazzo:2014rca}, in this case only slowly moving
quarkonium can be considered. This is so because for large
velocities the valid region of the saddle point approximation
strongly decreases and the absolute value of the imaginary part of
the potential vanishes.

\begin{figure}
\centerline{\includegraphics[width=3in]{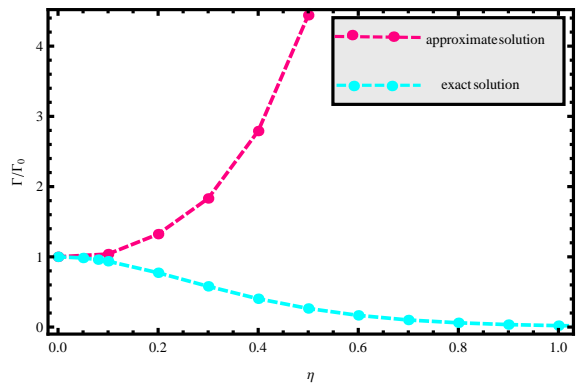}}%
\caption{ The thermal width versus the rapidity for two different
approaches. With using approximate solution, one finds the
increasing curve while exact solution gives us the decreasing curve.
}\label{N4TW}
\end{figure}

We have computed the thermal width from two different approaches in
Fig. \ref{N4TW}. It is clearly seen that the final results are not
the same. In the approximate solution, the width is increasing by
increasing the velocity of the quarkonium. We checked this behavior
for larger velocities and found that the width is monotonically
increasing. In the case of exact solution for imaginary potential,
as it is clear, the maximum value of $\G$ occurs for small velocity
of heavy quarkonium and one finds that the width vanishes for
ultra-high velocity. This is the case for a weakly coupled QCD
plasma in \cite{Escobedo:2013tca}. As it was pointed out, the case
corresponding to large quarkonium velocities would require to go
beyond the saddle point approximation. However, we see the different
behavior of the two approaches occurs at small velocities. These two
different observations can be found in a weakly coupled plasma in
\cite{Song:2007gm,Escobedo:2013tca}.

The approximate solution for a moving heavy quarkonium has been
followed in \cite{Ali-Akbari:2014vpa} and the results compared with
the weakly coupled calculations in \cite{Song:2007gm}. It is found
that increasing of velocity leads to higher decay rate for the
quarkoniums.

Regarding the poor assumption in the case of approximate solution,
we follow the exact solution to compute the width. This approach has
been followed also in the static case in
\cite{Fadafan:2013bva,Finazzo:2013rqy}. To compute the width for
$\Upsilon$, we assumed $m_Q=4.7GeV, \l=9, R=1$. For example at
$T=0.3 GeV$ and without any corrections, the width is $
\Gamma^{}=.487 MeV$. In the next plots for the width, one may
normalize the thermal widths in the presence of corrections to this
value.
\begin{figure}
\centerline{\includegraphics[width=3in]{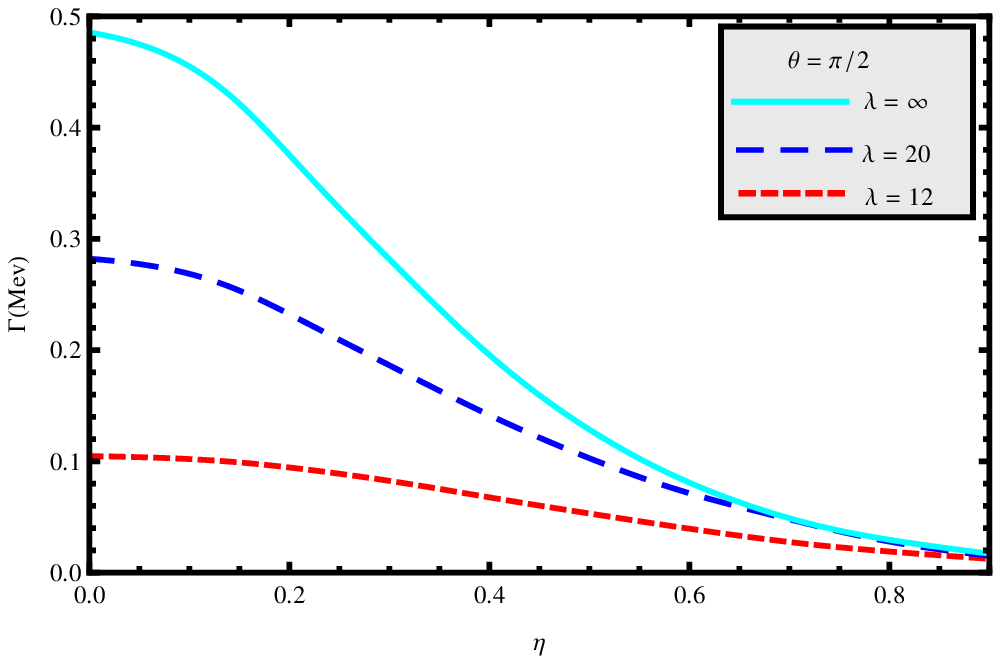}\,\,\includegraphics[width=3in]{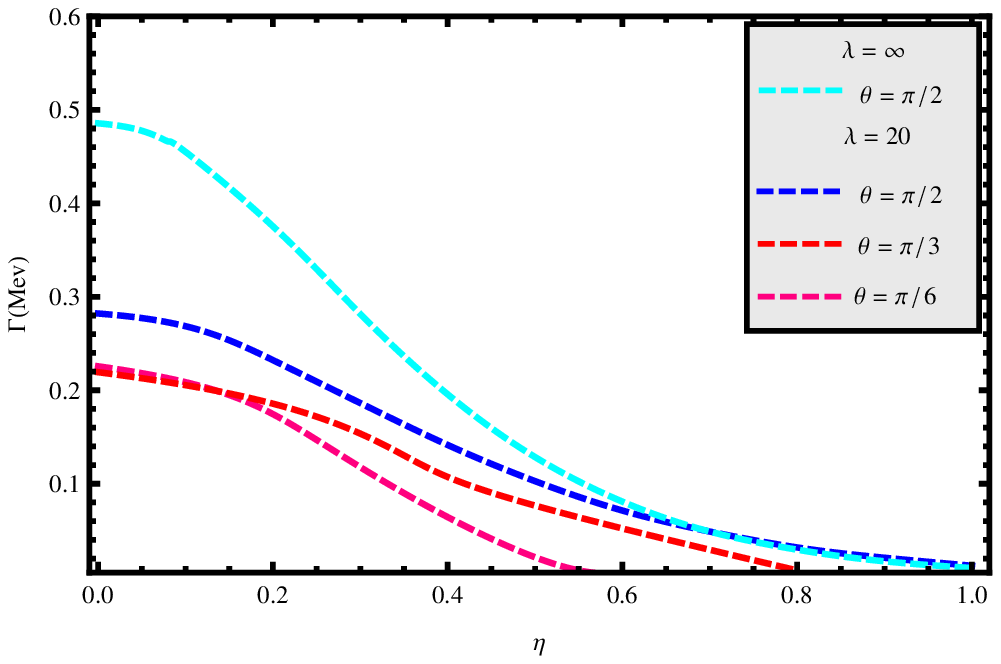}}%
\caption{ The thermal width versus the rapidity with finite coupling
corrections. In this figure $\l={\infty}$ means that we have not
considered any correction in the theory. Left: From top to down
$\l=\infty, 20, 12$ and $\theta=\pi/2$. Right: From top to down for
each curve, we assume $(\l=\infty, \theta=\pi/2)$, $(\l=20,
\theta=\pi/2)$,$(\l=20, \theta=\pi/3)$, $(\l=20, \theta=\pi/6)$.
}\label{R4TW10}
\end{figure}%
\begin{figure}
\centerline{\includegraphics[width=3in]{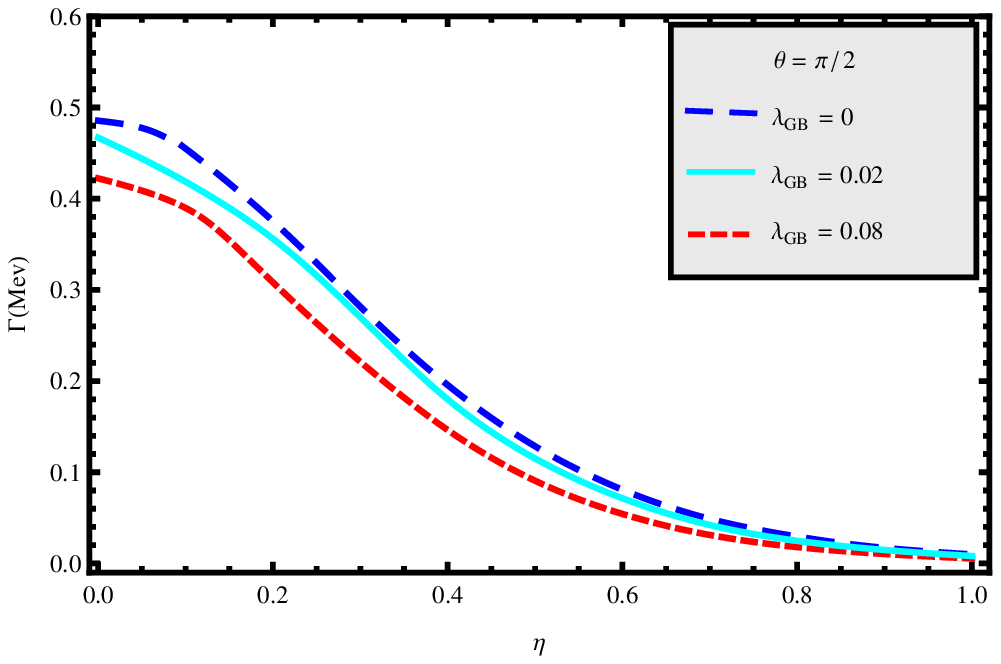}\,\,\includegraphics[width=3in]{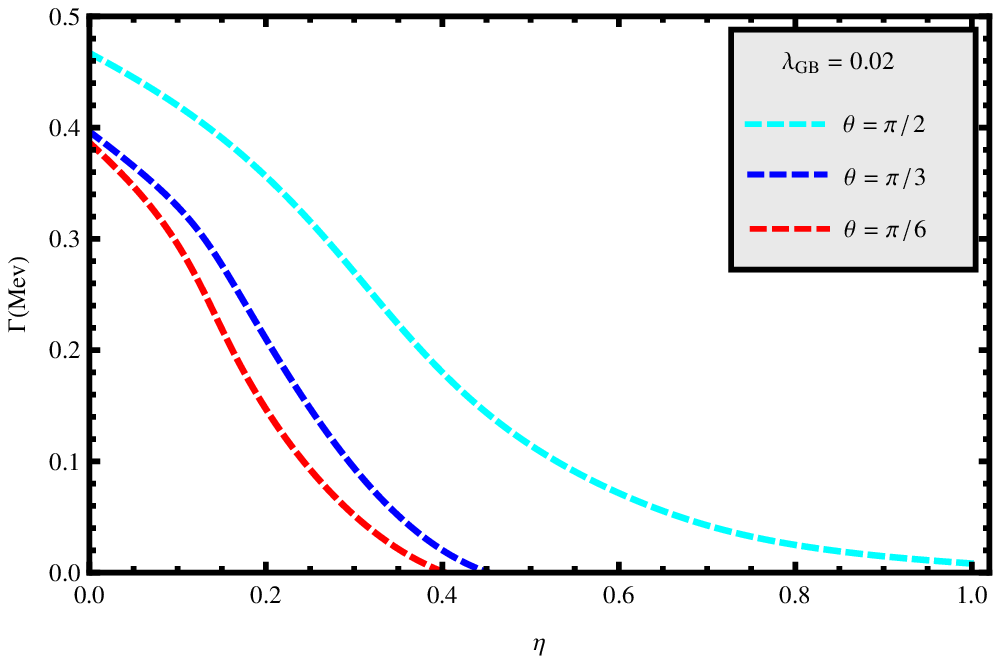}}%
\caption{ The thermal width versus the rapidity by considering the
Gauss$-$Bonnet corrections. In the left plot, we fixed the angle as
$(\theta=\pi/2)$ and from top to down $\lambda_{GB}$= 0, 0.02, 0.08.
In the right plot we fixed the Gauss$-$Bonnet coupling
$\l_{GB}=0.02$ and change the angle from top to down as
$(\theta=\pi/2,\pi/3,\pi/6)$. }\label{R4TW9}
\end{figure}

\subsection{Thermal width at finite coupling}
Now we plot the thermal width in the presence of $\mathcal{R}^4$ and
Gauss$-$Bonnet corrections in Fig. \ref{R4TW10} and Fig.
\ref{R4TW9}, respectively. In Fig. \ref{R4TW10}, we plotted the
thermal width versus the rapidity with the finite coupling
corrections. In this figure, $\l={\infty}$ means that we have not
considered any correction in the theory. By turning on the
corrections, $\l$ decreases and the width also reduces. We have
considered different angles for moving quarkonium to the wind in the
right plot. One confirms that decreasing angle leads to decreasing
the width.  In Fig. \ref{R4TW9}, we considered Gauss$-$Bonnet
quadratic curvature corrections to the AdS geometry. We fixed the
angle $(\theta=\pi/2)$ and the Gauss$-$Bonnet coupling
$(\l_{GB}=0.02)$ in the left and right plot, respectively. As it is
clear, in the left plot from top to down $\lambda_{GB}$= 0, 0.02,
0.08 and in the right plot from top to down
$(\theta=\pi/2,\pi/3,\pi/6)$. One concludes that at small rapidity
decreasing coupling ($\l$ or $\l_{GB}$) leads to decreasing thermal
width. More decreasing occurs in the case of $\mathcal{R}^4$
corrections. Thus one finds that the thermal width of static and
moving quarkonium decreasing with considering the higher derivative
corrections.

\section{Conclusion}
The main mechanism responsible for the suppression of quarkoniums is
proposed to be color screening \cite{Matsui:1986dk}. Recent studies
show a more important reason than screening, which is the existence
of an imaginary part of the potential. In this paper, we studied the
effects of finite 't Hooft coupling corrections on the imaginary
potential and the thermal width of a moving heavy quarkonium from
holography. An understanding of how these quantities change by these
corrections may be essential for theoretical predictions in
perturbative QCD \cite{Brambilla:2013dpa,Brambilla:2011sg}.

We discussed the limitations to the calculations of the imaginary
potential and the thermal width in our method. It was explained that
the valid region of the quark$-$antiquark distance for the onset of
the imaginary part of the potential depends on the velocity of
quarkonium. Also two different approaches for finding the thermal
width have been studied in Fig. \ref{N4TW}. Regarding the poor
assumption in the case of approximate solution, we followed the
exact solution to compute the width. As a result, one may conclude
that the width vanishes for larger velocities of quarkonium.

Comparing with the static quarkonium \cite{Fadafan:2013coa}, we find
qualitatively the same results by considering higher derivative
corrections. Briefly, in the presence of higher derivative
corrections the following is found.

\begin{itemize}%
\item{The absolute value of the imaginary part of the potential
of a moving or static heavy quarkonium becomes smaller.  }%
\item{The
thermal width of static or moving quarkonium decreasing with
considering the higher derivative corrections.}%
\end{itemize}%

Our results for the thermal width is similar to analogous
calculations in perturbative QCD . It was argued that at strong
coupling for a quarkonium with a very heavy constituent mass, the
thermal width can be ignored \cite{Lain1,Lain2} which is in good
agreement with our result. It is also interesting that in the ultra
relativistic limit such corrections are not so important. It has
been shown also that Gauss$-$Bonnet corrections can increase the
nuclear modification factor $R_{AA}$, which leads decreasing the
width \cite{Ficnar:2013qxa,Ficnar:2012yu}. We found the same result,
i.e. by considering the corrections the thermal width becomes
effectively smaller.

Including other phenomenological parameters controlling the plasma
properties would be interesting. It will be also very interesting to
investigate the thermal width of a quarkonia in more realistic
holographic backgrounds, such as in \cite{G1} and
\cite{Lee:2013oya}.

\section*{Acknowledgment}
We would like to thank J. Noronha and S. I. Finazzo for very useful
discussions.



\end{document}